\documentclass[showpacs,preprintnumbers,amsmath,amssymb,prl]{revtex4}

\usepackage{graphicx}
\usepackage{dcolumn}
\usepackage{bm}

\begin{document}

\title{Multiple-time correlation functions for non-Markovian interaction: Beyond the Quantum Regression Theorem}
\author{Daniel Alonso$^{\dag}$ and In\'es de Vega$^{\ddag}$}
\affiliation{$^{\dag}$Departamento de F\'{\i}sica Fundamental
y Experimental, Electr\'onica y Sistemas. Universidad de La Laguna,
La Laguna 38203, Tenerife, Spain \\ and $^{\ddag}$Departamento de F\'{\i}sica 
Fundamental II, Universidad de La Laguna, La Laguna 38203, Tenerife, Spain}

\begin{abstract}
Multiple time correlation functions are found in the dynamical description of different phenomena. They encode and describe the fluctuations of the dynamical variables of a system. In this paper we formulate a theory of non-Markovian multiple-time correlation functions (MTCF) for a wide class of systems.
We derive the dynamical equation of the {\it reduced propagator}, an object that evolve state vectors of the system conditioned to the dynamics of its environment, which is not necessarily at the vacuum state at the initial time. Such reduced propagator is the essential piece to obtain multiple-time correlation functions. An average over the different environmental histories of the reduced propagator permits us to obtain the evolution equations of the multiple-time correlation functions. We also study the evolution of MTCF within the weak coupling limit and it is shown that the multiple-time correlation function of some observables satisfy the Quantum Regression Theorem (QRT), whereas other correlations do not. We set the conditions under which the correlations satisfy the QRT. We illustrate the theory in two different cases; first, solving an exact model for which the MTCF are explicitly given, and second, presenting the results of a numerical integration for a system coupled with a dissipative environment through a non-diagonal interaction.
\pacs{3.65 Yz, 42.50 Lc}
\end{abstract}
\maketitle
{\sc{Introduction and motivation.}} Many research contexts are focused on the dynamics of a system $({\cal S})$ that is affected by an environment $({\cal B })$ from which it cannot be considered  $isolated$. Examples of such situations are encountered in statistical physics, condensed matter and quantum optics. We found a concrete example in the description of the dynamics of an atom $({\cal S })$ immersed in an electromagnetic field $({\cal B })$ \cite{carmichael1999,weiss1999}. 

In some circumstances, the analysis of the dynamics of the system is done using the expectation values of its observables over state vectors of the whole system, and then averaging over the environmental degrees of freedom. However, in some other situations, like when studying the response of a system to an external EM field, some additional information is needed. In particular, for the analysis of the spectroscopic properties of a system some {\it multiple-time correlation function} (MTCF) has to be computed, usually a two-time correlation function.

The dynamics of the system ${\cal S}$ is usually described through its reduced density operator. Such operator verifies some master equation that in the Markovian case is of Lindblad type \cite{lindblad1976,vankampen1981,carmichael1993,carmichael1999,zwanzig2001,gardiner2004}. Complementary to the master equation approach, a series of Monte-Carlo type of approaches based on the so called {\it stochastic Schr\"odinger equations} \cite{gisintot,carmichael1999,vankampen1981,diosi1995,plenio1998} have been developed in the last decade. In such schemes, the dynamics of system state vectors is integrated, and after an average is made over many realizations of environment $histories$ that eventually are understood as a $noise$ and takes into account the environment influence on ${\cal S}$. In the non-Markovian case, within the context of nuclear magnetic resonance, the Redfield master equation was developed \cite{redfield1957,redfield1965}. On the other hand, several non-Markovian stochastic Scr\"odinger equations have been established and studied very recently for systems influenced by a structured environment \cite{strunz1996,diosi1997,diosi1998,gaspard1999a,strunz1999b,cresser2000,gambetta2002}. 
In the Markovian case, an important tool to compute time correlation functions is the Quantum Regression Theorem (QRT) \cite{lax1963,carmichael1999,gardiner2004}. The theory of stochastic Schr\"odinger equations, initially elaborated to compute the expectation values of system observables, has been extended by many groups \cite{gisin1993a,brun1996,breuer1997,carmichael1999} to calculate multiple-time correlation functions for the Markovian case. In addition, such stochastic methods agree with the results expected from the QRT. It is then natural to develop an equivalent theory within the  stochastic Schr\"odinger equations approach, of multiple-time correlation functions for {\it non-Markovian} interactions. Interest arises particularly in those cases where the QRT {\it is not valid}.

Along with the theoretical interest, there is a practical need for a theory of non-Markovian MTCFs, in order to apply them to systems with experimental interest. Because of its potential applications, Photonic Band Gap (PBG) materials constitute an interesting example, in which the electromagnetic field displays a band-gap structure \cite{john1987,Yablonovitch1987}. As a consequence, the correlation function of the field is highly non-Markovian, and therefore an atom in interaction with such field will display non-Markovian dynamics. In this context, there are several works devoted to investigate the dynamics of few-level atoms in PGB-materials \cite{john1987,john1994,woldeyohannes2003}. For them, non-Markovian stochastic Schr\"odinger equations have been successfully applied \cite{devega2004}. 
Therefore, the theory of MTCF we intend to derive in this letter is important from both experimental and theoretical points of view.


{\sc{Multiple-time correlation functions.}} A frequently used Hamiltonian model in the study of the dynamics of $\cal{S}$ with Hamiltonian $H_S$, in interaction with $\cal{B}$ is 
\begin{eqnarray}
\label{Hamiltonian}
&&H=H_S+L B^\dagger+L^\dagger B +H_B
= \cr &&H_S+\lambda \bigg( L \sum_{n} g_n a_n^\dagger + L^\dagger \sum_{n} g_n^* a_n \bigg) + \sum_n \omega_n a_n^\dagger a_n,
\end{eqnarray}
where the operator $L$ acts on the Hilbert space of the system, $a_n,a_n^\dagger$ are the annihilation and creation operators on the environment Hilbert space and $\lambda$ is a suitable perturbation parameter that eventually can be taken equal to one. The $g_n's$ are the coupling constants and the $\omega_n's$ are the frequencies of the harmonic oscillators that constitute the environment \cite{caldeira1983}.

We are interested in the evaluation of {\it N-time correlation functions}, defined for a set of observables $\{A_1(t_1),\cdots,A_N(t_N)\}={\bf A({\bf t})}$ in Heisenberg representation as
\begin{equation}
\label{mtcf}
C_{{\bf A}}({\bf t}|\Psi_0)\equiv
\langle \Psi_0 | A_1(t_1) \cdots A_N(t_N)|\Psi_0 \rangle,
\end{equation}
with $t_1>t_2>\cdots >t_N$ and ${\bf t}=\{ t_1,\cdots,t_N \}$.
The initial state of the full system is taken as the tensor product of a system state $|\psi_0\rangle$ and the environment state $|z_0\rangle$, $i.e.$ $|\Psi_0\rangle= |\psi_0\rangle|z_0\rangle$. 

In the partial interaction picture with respect to the environment, the N-time correlation function is defined as $C_{{\bf A}}({\bf t}|\Psi_0)= \langle \Psi_0|\prod_{i=1}^N {{\cal U}}^{-1}_I(t_i,0)A_i{{\cal U}}_I(t_i,0)|\Psi_0 \rangle,$
where ${{\cal U}}_I$ is the evolution operator of the system in the interaction picture. A suitable basis to treat the environment ${\cal B}$ is a coherent state basis, $|z_1,z_2,\cdots,z_n,\cdots \rangle=|z\rangle$ in the Bargmann representation \cite{carmichael1999,strunz2001a}. In such basis the resolution of the identity is given by ${\sl 1}=\int d\mu(z) |z\rangle \langle z|$ with $d\mu(z)=\prod_{i=1}^N (d^2z_i  \exp(-|z_i|^2)/\pi)$, and when inserted in the definition of $C_{{\bf A}}({\bf t}|\Psi_0)$ it follows
\begin{eqnarray}
\label{eqn2}
C_{{\bf A}}({\bf t}|\Psi_0)=
\int d\mu(z)\langle \psi_0|
G^{-1}(0,1) \prod_{i=1}^{N} A_i G(i,i+1)
|\psi_0 \rangle,
\end{eqnarray}
with $t_0=0$, $t_{N+1}=0$ and $z_{N+1}=z_0$. We have introduced the {\it reduced propagators} $G(i,i+1)\equiv G(z_i^* z_{i+1}|t_it_{i+1})=\langle z_i | {{\cal U}}_I(t_i,t_{i+1})| z_{i+1} \rangle$, which act on the system Hilbert space, giving the evolution of system state vectors from $t_{i+1}$ to $t_i$, conditioned that in the same time interval the environment coordinates go from $z_{i+1}$ to $z_i$. Once their time evolution is solved, the time-correlation function (\ref{eqn2}) can be obtained. Therefore, to proceed further we need to derive the equation of motion of the reduced propagator $G(i,i+1)$, by considering its time derivative with respect to $t_i$. Taking into account the fact that the evolution operator ${{\cal U}}_I$ satisfies the Schr\"odinger equation in the partial interaction picture, and after some manipulations, we arrive to the equation
\begin{eqnarray}
\label{eqn12}
\frac{\partial G(i,i+1)}{\partial t_i}=
\big(-i H_S+L z^*_{i,t_i}- L^\dagger z_{i+1,t_i}\big)G(i,i+1)\cr
-L^\dagger \int_{t_{i+1}}^{t_i} d\tau 
\alpha(t_i-\tau)
\langle z_i | {{\cal U}}_I(t_i,t_{i+1})L(\tau,t_{i+1})|z_{i+1} \rangle,
\end{eqnarray}
with $L(t',t)=e^{iH_Bt}e^{-iH(t-t')}Le^{iH(t-t')}e^{-iH_Bt'}$,$z_{i,t}=i\sum_n g_n z_{i,n}e^{i \omega_n t}$ and $\alpha(t-\tau)=\sum_n |g_n|^2 e^{-i \omega_n (t-\tau)}$.
The function $z_{i,t}$ is a sum over time dependent coherent states and $\alpha(t-s)$ is its time autocorrelation function, as it can be verified by computing the average $M[z_{i,t}z^*_{i,s}]$ with respect to the measure $d\mu(z)$. When integrating equation (\ref{eqn12}), we still have the problem that it is not possible to compute exactly the matrix element$\langle z_i | {{\cal U}}_I(t_i,t_{i+1})L(\tau,t_{i+1})|z_{i+1} \rangle$, and to express it as a function of the reduced propagator. This would bring (\ref{eqn12}) to an explicit equation for the reduced propagator. Since only in very exceptional cases exact solutions can be obtained, some approximate scheme has to be taken.
One possible way is to treat $L(\tau,t_{i+1})$ in the weak coupling limit. Else, sometimes it is possible to assume that $\langle z_i | {{\cal U}}_I(t_i,t_{i+1})L(\tau,t_{i+1})|z_{i+1} \rangle= O(z_{i+1} z_{i},t_{i+1},\tau)G(i,i+1) $, where the operator $O$ has to be constructed \cite{yu1999}. In this case we have
\begin{eqnarray}
\label{eqn12b}
&&\frac{\partial G(i,i+1)}{\partial t_i}=
\bigg(-i H_S+L z^*_{i,t_i}- L^\dagger z_{i+1,t_i}\cr
&&-L^\dagger \int_{t_{i+1}}^{t_i} d\tau 
\alpha(t_i-\tau) O(z_{i+1} z_{i},t_{i+1},\tau)\bigg)G(i,i+1).
\end{eqnarray}
The equation (\ref{eqn12}) or its approximate versions, in particular equation (\ref{eqn12b}), depends on two time dependent functions, $z^*_{i,t_i}$ and $z^*_{i+1,t_i}$, that take into account the ``history'' of the environment and lead to a conditioned dynamics of the system with respect to the environment dynamics. They constitute one of the results of this letter and they  are the starting point to compute the MTCF in the non-Markovian case. The integration of the equations for the reduced propagators along with their initial conditions, $G(i,i+1)=\exp{(z_i^* z_{i+1})}$ , leads to the evaluation of the $N$-time correlation functions previously defined. In addition, let us comment that since the equation for the reduced propagator (\ref{eqn12}) is made for an initial state of the environment different form the vacuum, this equation permits us to evaluate expectation values and correlation functions of system observables with  \emph{more general initial conditions that the one usually taken}, $i.e.$ $|\Psi_0 \rangle = |\psi_0 \rangle |\hbox{vacuum}\rangle$, but we will not report those results here.

In principle, and under the conditions we have considered, the non-Markovian multiple-time correlation functions derived in this section are rather general. We shall look closer to the time correlation functions we have derived, showing how they can be related to the Quantum Regression Theorem which applies for the Markovian case, and we will show a couple of examples.

{\sc{Beyond the Quantum Regression Theorem. Weak-coupling limit.}} Once we have the multiple time correlation functions, we may compute them directly from the stochastic method. Nonetheless, this may turn to be an expensive strategy from the numerical point of view, what is specially true when the number of environmental degrees of freedoms needed to correctly describe its correlation function is large. For those cases, we present in this section a set of coupled differential equations in which the stochastic average has been done analitically, and which evolve the non-Markovian two-time correlations up to second order in a convenient perturbative parameter $\lambda$. 

The method we will follow consists in deriving the reduced two-time correlation $ \langle\psi_{0}\mid G^{I \dagger}(z_1 0|t' 0) V_{t'}A G^{I}(z_1^* z_2|t' t)V_t BG_{t,0}(z^*_{2} 0|t 0)\mid\psi_{0}\rangle$ with respect to $t'$, and then performing analitically the average over the variables $z_1$ and $z_2$. In order to do that, it is necessary to use a perturbative expansion of the propagators in the interaction picture with respect to the system, which is used because it leads to more handable expressions. We then arrive to the following set of differential equations for the two-time correlation functions up to ${\mathcal{O}}(\lambda^3)$
\begin{eqnarray}
&&\frac{d}{dt'}\langle\Psi_{0}\mid A(t')B(t)\mid\Psi_{0}\rangle= \big \langle\Psi_{0}\big| \Big( i \left\{[H_s ,A]\right\}(t')B(t)\nonumber\\
&+& \int^{t'}_0 d\tau \alpha^*(t'-\tau)\left\{V_{\tau-t'} L^{\dagger}[A,L]\right\}(t')B(t)\nonumber\\
&+&\int^{t'}_0 d\tau \alpha(t'-\tau) \left\{[L^{\dagger},A]V_{\tau-t'} L \right\}(t')B(t)\nonumber\\
&+&\int_0^t d\tau \alpha(t'-\tau) \left\{[L^{\dagger}, A]\right\}(t')\left\{[B,V_{\tau-t} L]\right\}(t)\Big) \big| \Psi_{0}\big\rangle.
\label{total3}
\end{eqnarray}
In this expression, the time dependencies of operators are now in the total interaction image evolution operator. We also  denote $V_{t'}L \equiv \exp\{i{\mathcal L}_S t'\}L=\exp{(iH_S t')}L \exp{(-iH_S t')}$, where $V_{t'}=\exp\{i{\mathcal L}_S t'\}$ is the free system Liouville operator, acting in the two sides of the immediately contiguous system operator. It can be checked that only when the last term of (\ref{total3}) vanishes, i.e. when $[L^{\dagger},A]=0$ or $[B,V_{\tau-t} L]=0$, the Quantum Regression Theorem applies. Notice also that this term is zero in the Markovian case, since the corresponding correlation function $\alpha(t'-\tau)=\Gamma \delta(t'-\tau)$ vanishes in the domain of integration from $0$ to $t$. In summary, the previous equations leads to the computation of the MTCF and they contain the conditions under which the QRT remains valid in the weak coupling limit. Equation (\ref{total3}) is another result of this letter.

{\sc{A solvable example.}} To illustrate the theory proposed in this letter, we shall apply it to a simple solvable model described by the hamiltonian (\ref{Hamiltonian}) with $L=\sigma_z$ and $H_S=\frac{\omega}{2}\sigma_z$. This model describes the dynamics of system state vectors towards one of the eigenstates of the system Hamiltonian. Notice that in this case $[H_S,L]=0$ and then $O=L$ (see equation \ref{eqn12b}). 

Let us consider the two-time correlation of $A=\{\{0,\alpha\},\{\beta,0\}\}$ and $B=\{\{1,0\},\{-1,0\}\}=\sigma_z$. For an initial system state $|\psi_0 \rangle= \psi_{01} |\, + \, \rangle+\psi_{02}|-\rangle$ the result is
\begin{eqnarray}
\label{model9}
C_{AB}(t't|\Psi_0)= e^{-2 I_{00}^{t't'}(\alpha)}
\big\{\beta  \psi_{02}^* \psi_{01} e^{-i \omega t'}
-\alpha \psi_{01}^* \psi_{02} e^{i \omega t'}
\big\},
\end{eqnarray}
with the definition $I_{ac}^{bd}(\alpha)\equiv \int_a^{b} d\tau \int_c^{d} ds \alpha(\tau-s)$. For the case in which $A=B=\sigma_z$, we have $C_{\sigma_z \sigma_z}=1$. Another type of two-time correlation function is the one corresponding to the observables $A=\{\{0,\alpha\},\{\beta,0\}\}$ and $B=\{\{0,\alpha '\},\{\beta ',0\}\}$ given by
\begin{equation}
\label{model14}
C_{AB}(t't)= e^{\tilde{D}(t't)} \big\{ \alpha \beta' |\psi_{01}|^2 e^{i \omega (t'-t)}+
\alpha' \beta |\psi_{02}|^2 e^{-i \omega (t'-t)}
\big\},
\end{equation}
with $\tilde D(t't)=I_{00}^{t'\tau}(\alpha^*)+I_{tt}^{t'\tau}(\alpha)+I_{00}^{t\tau}(\alpha)
+I_{0t}^{t't'}(\alpha)-I_{t0}^{t't}(\alpha)-I_{00}^{t't}(\alpha)$.
In the first case, since $[L,B]=0$ the last term of equation (\ref{total3}) is zero, and the QRT is valid; meanwhile, in the second type of correlation functions neither $[L,B]=0$ nor $[L^{\dagger},A]=0$ and therefore the QRT is not expected to hold. This can be verified by direct substitution of the two-time correlation functions obtained into (\ref{total3}), giving further support to the equations proposed.

\begin{figure}[htbp]
\centerline{\includegraphics[width=0.3\textwidth]{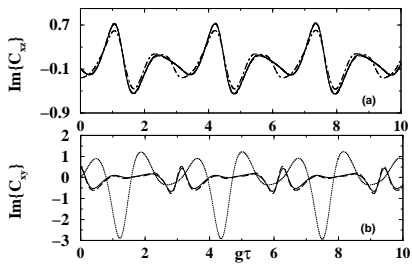}}
\caption{Figure (a) represents the imaginary part of $C_{\sigma_x \sigma_z}$, the solid line represents the analytical result (\ref{model9}), that in {\it this case} coincides the Quantum Regression Theorem result. Dotted an dashed lines represent the stochastic result with an average over $10^2$ and $10^5$ trajectories respectively. Figure (b) represents the imaginary part of $C_{\sigma_x \sigma_y}$. Comparing the result of the QRT (dotted line), to the exact result given by (\ref{model14}) (solid line), it is clear that QRT is not valid in this case. The dot-dashed and the long-dashed lines are obtained averaging over $10^2$ and $10^4$ trajectories respectively.}
\label{fig11}
\end{figure}

As an illustration, the figure (\ref{fig11}) shows two-time correlation functions of the system $C_{\sigma_x \sigma_z}$ and $C_{\sigma_x \sigma_y}$, with two oscillators in the environment with parameters $g_1,g_2=g=1$ and $\omega_1=6, \omega_2=2$. The initial system state taken in all computations in this letter is $|\Psi_0 \rangle= | \psi_0 \rangle |\hbox{vacuum}\rangle$ with $|\psi_0\rangle= \big((1+2 i)|+\rangle + (1+i)|-\rangle \big)/\sqrt{7}$. 
It is clear from the figures (c) and (d) that the QRT does not apply for $C_{\sigma_x \sigma_y}$, since $[L,B]\neq 0$ and $[L^{\dagger},A]\neq 0$.

\begin{figure}[htbp]
\centerline{\includegraphics[width=0.3\textwidth]{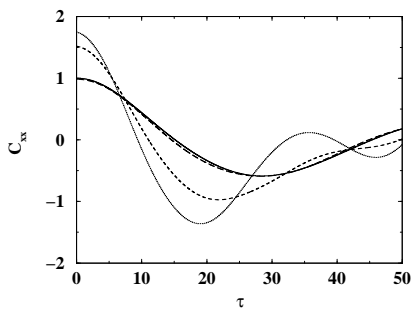}}
\caption{Two time correlation $C_{\sigma_x \sigma_x}(t',t)$ for the coupling $L=\sigma_{12}$, and the Fourier series of the exponential correlation function with $\nu=8$ oscillators. The parameters are: $\omega =0.1$, $\Gamma=1$, perturbative parameter $g=0.2$, recurrence time $T=40$, and initial time for the correlation $t=1$. Solid line represents the solution of the system (\ref{total3}), whereas dotted, dashed and long dashed lines gives respectively the result of the stochastic method for $\kappa=5\times 10^5, 10^6, \, \hbox{and} \, 6 \times 10^7$ trajectories. An increasing accordance with the system curve (equation \ref{total3}) is observed as the number of trajectories grows.}
\label{patatin}
\end{figure}

\begin{figure}[htbp]
\centerline{\includegraphics[width=0.3\textwidth]{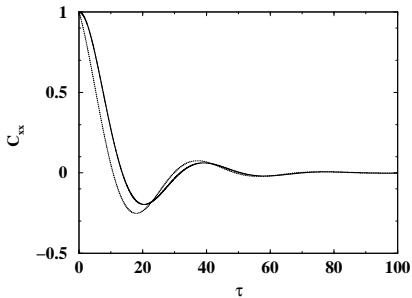}}
\caption{Two time correlation $C_{xx}=C_{\sigma_x \sigma_x}$ (see text) The parameters are: $\omega =0.1$, $\Gamma=1$, perturbative parameter $g=0.4$, and initial time for the correlation $t=10$. Solid line represents the solution of the system (\ref{total3}), and dotted line gives the result expected with the QRT. Because the last term in (\ref{total3}) is non zero, both results are different from each other, and the QRT is not valid.}
\label{diss2}
\end{figure}

{\sc{An example of dissipative system}}. Let us now apply the theory derived in this paper for the case $L=\sigma_{12}$, and for a dissipative interaction. Within the perturbative approximation, the operator $O(z,t,\tau)$ can be replaced by its zero order perturbative expansion, $V_{\tau-t}L=\sigma_{12}\exp{\{i\omega (t-\tau)\}}$, where $\omega$ is the system rotating frequency. 
We propose the following correlation function, $\alpha(t-\tau)=\sum_{m=-\nu/2}^{\nu/2} C(m)e^{-i\pi m (t-\tau)/T}$, with the coefficients,$C(m)=\frac{1}{2T}\int^{T}_{-T} dt \alpha(t)e^{i \pi mt/T}$, which represents the Fourier series of the function $(\Gamma/2) \exp{\{\Gamma|t-\tau|\}}$. In those equations, $T$ is the time window in which the correlation function is expanded in the series. The more members we add in the former sum, the closer the solution is to the exponential decaying correlation function, and the larger we can fix the recurrence time $T$. The decaying behaviour is displayed in Figure (\ref{patatin}), representingthe correlation $C_{\sigma_x \sigma_x}$ as a function of $t'-t=\tau$. In this figure, we compare the evolution given by (\ref{total3}) to the ensemble averaged stochastic evolution for different number of trajectories.
We also consider the correlation function $\alpha(t-\tau)=(\Gamma/2) \exp{\{-\Gamma|t-\tau|\}}$, in order to study the validity of the QRT for $C_{\sigma_x \sigma_x}$ (see Figure \ref{diss2}).

In summary, we derive in this letter a theory for non-Markovian multiple-time correlation functions from stochastic Schr\"odinger equations. The starting point of the theory is the evolution equation for the reduced propagator, that evolves vectors in the Hilbert space of the system ${\mathcal S}$ conditioned to the dynamics of the environment. Remarkably, such equation depends on two different histories of the bath. With the reduced propagator, the multiple time correlation functions are formally obtained. Furthermore we have derived, in the weak-coupling limit, the set of coupled differential equations that satisfy the two-time correlation functions. This set of equations is a generalisation of the Quantum Regression Theorem, and they show clearly the cases in which such theorem is not longer valid. We have verified the theory by applying it to two cases. For a solvable model, we computed the two-time correlation functions explicitly, displaying a case in which the QRT is fulfilled and a case in which it is not. For a case in which the system has a non-diagonal interaction with the bath, and the bath has a decaying exponential function, we have numerically integrated the multiple-time correlation functions. Following the procedures here established, higher order time correlation functions might be calculated. We believe that this work is relevant, aside from its intrinsic theoretical interest, in the description of the dynamics of small systems, such as atoms immersed in photonic crystals as well as other situations where non-Markovian effects are relevant.

{\sc Acknowledgements} We thank G.C. Hegerfeldt for his comments and the support of the Gobierno de Canarias (Spain) (PI2002/009) and Ministerio de Ciencia y Tecnolog\'\i a of Spain (BFM2001-3349). I. de Vega is financially supported by a Ministerio de Ciencia y Tecnolog\'\i a (AP2001-2226).
\bibliographystyle{apsrev}

\end{document}